\documentclass[sigconf,screen]{acmart}

\usepackage{booktabs} 
\usepackage{amsfonts}
\usepackage{amssymb}
\usepackage{xspace}
\usepackage{epstopdf}
\usepackage{enumitem}
\usepackage{algorithm}
\usepackage{algorithmic}
\usepackage{multirow}
\usepackage{balance}
\usepackage{booktabs}
\usepackage{array}
\usepackage{wrapfig}
\usepackage{pifont}
\usepackage{pbox}
\usepackage{geometry}
\usepackage{color}
\begin{document}

\def\x{{\mathbf x}}
\def\L{{\cal L}}
\def\x{{\mathbf x}}
\def\L{{\cal L}}
\def\eg{\textit{e.g.}}
\def\ie{\textit{i.e.}}
\def\Eg{\textit{E.g.}}
\def\etal{\textit{et al.}}
\def\etc{\textit{etc.}}
\def\nback{\textit{n}-back }
\setlength{\tabcolsep}{2pt}

\title{Engagement Estimation in Advertisement Videos with EEG}




\author{Sangeetha Balasubramanian}
\affiliation{%
  \institution{International Institute of Information Technology}
  \city{Hyderabad} 
  \country{India} 
}
\email{sangeetha.balasubramanian@students.iiit.ac.in}

\author{Shruti Shriya Gullapuram}
\affiliation{%
 \institution{University of Massachusetts Amherst}
  \city{Amherst} 
  \country{United States} 
}
\email{sgullapuram@umass.edu}

\author{Abhinav Shukla}
\affiliation{%
  \institution{International Institute of Information Technology}
  \city{Hyderabad} 
  \country{India} 
}
\email{abhinav.shukla@research.iiit.ac.in}

\renewcommand{\shortauthors}{S. Balasubramanian, S. S. Gullapuram, A. Shukla}

\begin{abstract}
Engagement is a vital metric in the advertising industry and its automatic estimation has huge commercial implications. This work presents a basic and simple framework for engagement estimation using EEG (electroencephalography) data specifically recorded while watching advertisement videos, and is meant to be a first step in a promising line of research. The system combines recent advances in low cost commercial Brain-Computer Interfaces with modeling user engagement in response to advertisement videos. We achieve an F1 score of nearly 0.7 for a binary classification of high and low values of self-reported engagement from multiple users. This study illustrates the possibility of seamless engagement measurement in the wild when interacting with media using a non invasive and readily available commercial EEG device. Performing engagement measurement via implicit tagging in this manner with a direct feedback from physiological signals, thus requiring no additional human effort, demonstrates a novel and potentially commercially relevant application in the area of advertisement video analysis.
\end{abstract}


\settopmatter{printacmref=false}
\renewcommand\footnotetextcopyrightpermission[1]{}
\pagestyle{plain}

\maketitle

\section{Introduction}~\label{sec:intro}
The popularity and versatility of social networks allows organizations to reach their chosen target audience and offer products and services that suit their constantly changing needs best. In this digital era, advertisements are the best way to do so. Emotion and engagement play a powerful tool in determining a person's intent to buy a product. Positive and negative emotions are used by advertisers to promote their product and drive users into buying them. Brands associated with positivity has been shown to increase engagement. However, negative emotion is also used to effectively garner consumer attention, where certain life choices are portrayed as beneficial and improving one's quality of life, while others are portrayed as harmful and potentially fatal. Advertisement \textit{valence} (pleasantness), \textit{arousal} (emotional intensity) and \textit{engagement} (emotional involvement) are key properties that play a major role in consumer attitudes associated with the advertised product ~\cite{Holbrook1984,Holbrook1987,Pham2013}. Analyzing advertisements in terms of emotional content can also help optimize user experience ~\cite{cavva}. 

This work expressly investigates the utility of physiological EEG (electroencephalography) signals in the task of estimating engagement in affective advertisement videos (evenly distributed over the emotional dimensions). 14 channel EEG data is acquired via the commercial \textit{Emotiv} headset while the test subjects are watching the advertisements in question, and then subsequently analyzed for engagement signatures. An implicit tagging based approach is used to estimate the engagement level, without requiring any additional effort from the end user.

In summary, we make the following contributions: 1) Our work is one of the first to correlate user engagement in advertisement videos with the recorded EEG signal 2) We demonstrate a simple yet effective way that a low cost commercial EEG sensor can be employed for engagement recognition in the wild

Our paper is organized as follows. Section \ref{RW} discusses related work and espouses the novelty of our work with respect to existing literature. Section \ref{ad_set} describes the ad dataset used in our experiments. Section \ref{EEGprotocol} shows the EEG data acquisition apparatus and protocol via the \textit{Emotiv} headset. \ref{ER} discusses experiments and key respective findings. Section \ref{CFW} concludes the paper and highlights important potential future implications and considerations.

\begin{figure}[t]
\includegraphics[width=\linewidth]{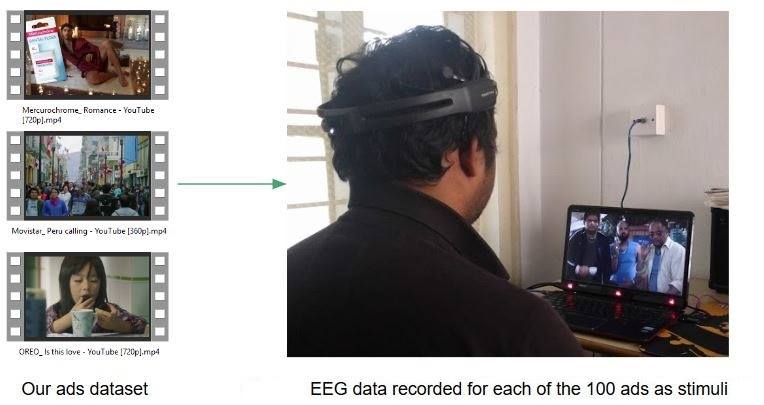}\hfill
\caption{\label{eeg} An EEG recording session with a subject}
\end{figure}

\section{Related Work}\label{RW}
To highlight the novelty of our work, we briefly review prior works
examining (i) Conversational Systems, (ii) Engagement in online content, and (iii) Online advertising.

\subsection{Conversational Systems}
Various multimodal features and machine learning algorithms are used to predict engagement or involvement in human-human  and human-robot conversations~\cite{yu2015ticktock,celiktutan2017multimodal}. Some work use more fine grained features, such as Oertel et al.~\cite{oertel2013gaze} used gaze features obtained by an eye tracker to model engagement in multiparty conversations. In more recent work~\cite{yu2015attention}, human behaviors, such as smiles and speech volume was used to coordinate with users' engagement on the fly via techniques such as adaptive conversational strategies and incremental speech production. 

\begin{figure}[t]
\includegraphics[width=0.8\linewidth]{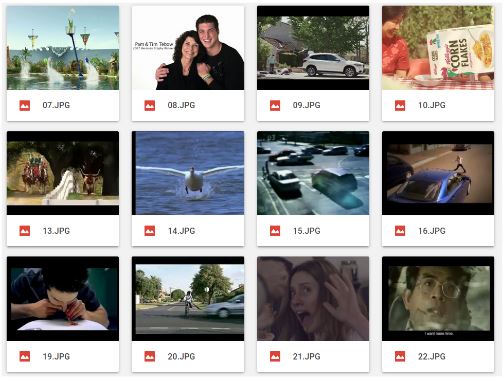}\hfill
\caption{\label{thumbnails} Representative thumbnail images from the videos in our dataset}
\end{figure}

\subsection{Engagement in online content}
Many researchers have analyzed engagement behavior towards web content.The line of work that measures video engagement is often around measuring engagement from the video popularity dynamics (number of views, likes, \etc) ~\cite{park2016data}. More recent work~\cite{wu2018beyond} focuses on measuring video engagement from public information, such as video context, topics, number of times a video is watched and channel, without observing any user reaction. While a lot of work is focused on engagement and attention in online web content and videos, there is very little work determining the engagement in advertisement videos.

\subsection{Online advertising}
Advertising is most effective when it elicits an emotional response and engages the viewer on a personal level. There has been extensive study in the context of display advertising ~\cite{azimi2012visual, rosales2012post} and sponsored search ~\cite{becker2009happens, sculley2009predicting, sodomka2013predictive}. A lot of work focused on predicting performance of an ad based on its click-through rate. Later efforts has been on  matching the web queries or pages to the context of the ad ~\cite{becker2009happens, becker2009context,lee2013semantic, oentaryo2014predicting}. More recent work ~\cite{barbieri2016improving} focuses on analyzing the level of user engagement with ads as the length of time a user spends after landing on an ad page. There have also been recent attempts to use EEG to mine affective attributes from data \cite{Shukla2017icmi} and use it for more emotionally relevant advertising. A very recent work also uses contextual and visual attention data from eye tracking to suggest emotionally driven design principles for advertisements \cite{Shukla2018context}.

However, we specifically study the potential contribution of the physiological EEG modality to engagement estimation. We use an emotional advertisement dataset curated by \cite{Shukla2017acm}, the details of which follow in the next section. These ads are then presented as stimuli to test subjects whose EEG signal data is recorded. The subjects also self-report engagement on a 5-point scale. Engagement estimation is achieved via EEG signals acquired via the wireless and wearable \textit{Emotiv} headset, which facilitates naturalistic user behavior and can be employed for large-scale engagement recognition from multimedia content.
 
\section{Advertisement Dataset Description}\label{ad_set}
This section presents details regarding the ad dataset used in this study..

We use an affective advertisement dataset curated by \cite{Shukla2017acm} for our study. Defining \textbf{\textit{valence}} as the degree of \textit{pleasantness}/\textit{unpleasantness} and \textbf{\textit{arousal}} as the \textit{intensity of emotional feeling}, five experts carefully compiled a dataset of 100, roughly 1-minute long commercial advertisements (ads). These ads are publicly available\footnote{On video hosting websites such as YouTube}. The authors of~\cite{Shukla2017acm} chose the ads based on consensus among five experts on valence and arousal labels (either \textit{high} (H)/\textit{low} (L)). High valence ads typically involved product promotions, while low valence ads were social messages depicting ill effects of smoking, alcohol and drug abuse.

\begin{figure}[t]
\includegraphics[width=0.8\linewidth]{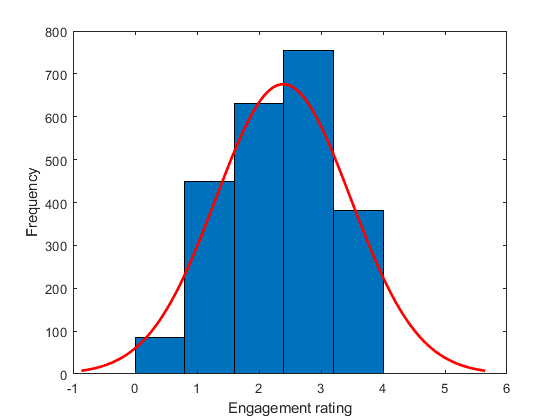}\hfill
\caption{\label{hist} Engagement rating distribution with Gaussian pdf overlay}
\end{figure}

\begin{table}[h]
\fontsize{10}{10}\selectfont
\renewcommand{\arraystretch}{1.5}
\caption{\label{Corr_Ratings} Mean correlations between self-rated attributes. Significant correlations ($p<0.05$) are denoted in bold.} \vspace{-.1in}
\centering
\begin{tabular}{l|lll}
                     & \textbf{Arousal} & \textbf{Valence} & \textbf{Engagement} \\ \hline \hline
\textbf{Arousal}     &  1    & -0.19    & \textbf{0.36}     \\
\textbf{Valence}     & ~    & ~1         & 0.11       \\
\textbf{Engagement}  & ~    & ~          & ~1         \\
\hline
\end{tabular}
\end{table}

To evaluate the effectiveness of these ads as affective control stimuli, \cite{Shukla2017acm} examined how consistently they could evoke target emotions across viewers. To this end, the ads were independently rated by 14 annotators for valence (valence) and arousal in \cite{Shukla2017acm}. All ads were rated on a 5-point scale, which ranged from -2 (\textit{very unpleasant}) to 2 (\textit{very pleasant}) for valence and 0 (\textit{calm}) to 4 (\textit{highly aroused}) for arousal.

In this work, we chose a similar setting for the rating of engagement. Defining \textbf{\textit{engagement}} as the \textit{emotional involvement or commitment} while viewing an audio-visual stimulus on a 5-point scale from 0 (\textit{least engaging}) to 4 (\textit{most engaging}) as shown in Fig.~\ref{eng_scale}, we employed 23 annotators for the engagement rating task, with each person viewing a variable number of ads. For each advertisement, we calculated the mean of ratings of all annotators and then thresholded this average by the grand average of all ratings of all ads to get a binary label that we treated as \textbf{ground truth} in all our engagement estimation experiments.

The distribution of engagement ratings (Fig.~\ref{hist}) is roughly uniform resulting in a Gaussian fit with large variance, with the mean observed at the median scale value of 2.

Pearson correlation was computed between the arousal, valence and engagement (see table \ref{Corr_Ratings}) dimensions by limiting the false discovery rate to within 5\%~\cite{benjamini1995controlling}. This procedure revealed a weak and insignificant negative correlation of 0.17, implying that ad arousal and valence scores were largely uncorrelated. However, we found that arousal and engagement had a statistically significant positive correlation of 0.36, which goes to show that advertisements with high affective activation tend to be more engaging.

\section{EEG Acquisition Protocol}\label{EEGprotocol}

\begin{figure}[t]
\includegraphics[width=0.7\linewidth]{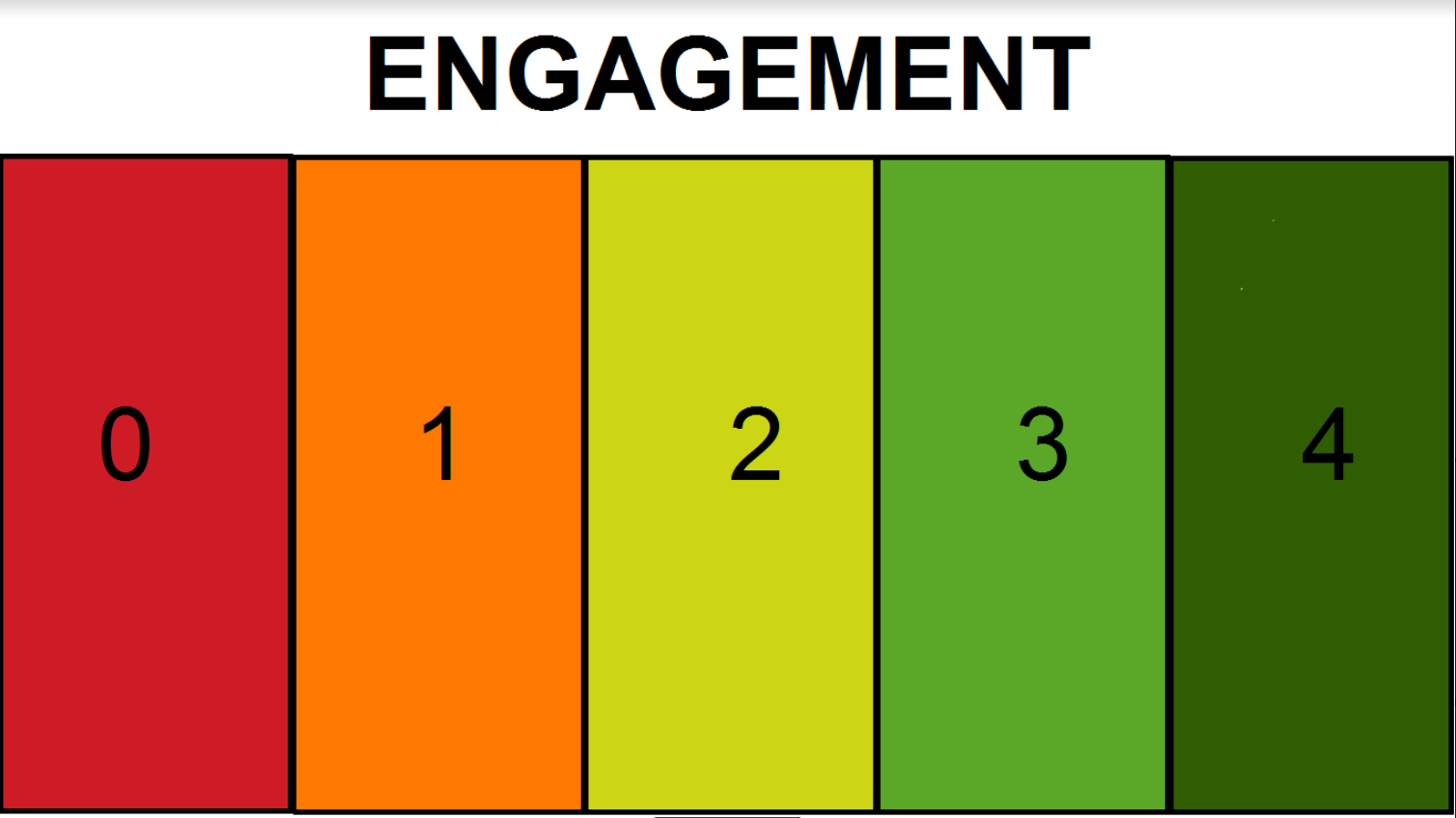}\hfill
\caption{\label{eng_scale} Rating prompt shown to the subject after each stimulus advertisement}
\end{figure}

\begin{figure}[t]
\includegraphics[width=0.8\linewidth]{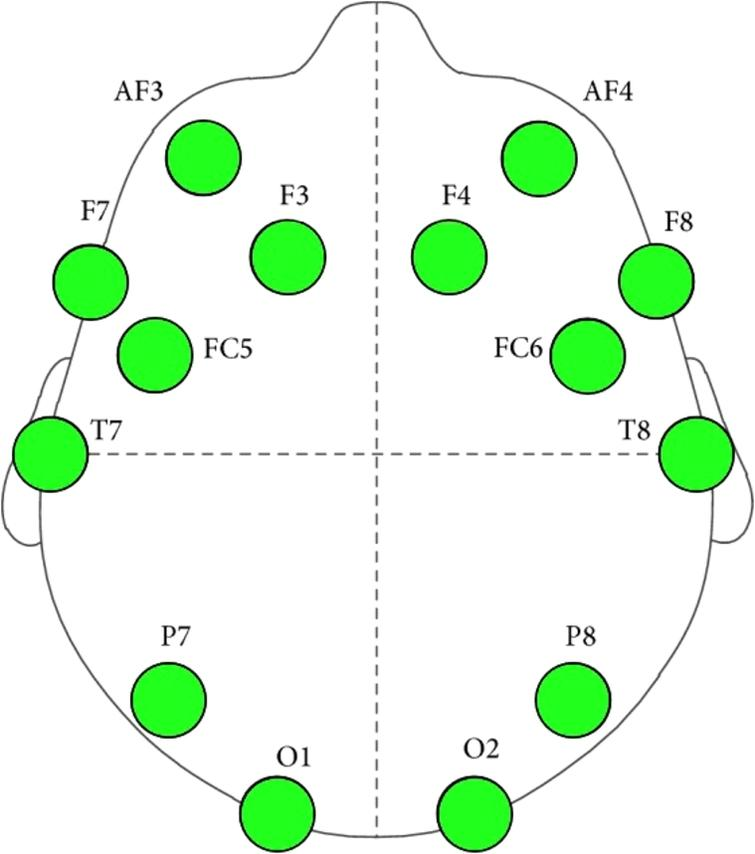}\hfill
\caption{\label{eeg_locs} The electrode locations of the 14 channel \textit{Emotiv} EEG headset}
\end{figure}

As the annotators rated the ads for engagement upon watching them, we acquired their Electroencephalogram (EEG) brain activations via the \textit{Emotiv} wireless headset as shown in Fig.~\ref{eeg}. To maximize attention and minimize fatigue during the rating task, these raters took a break after every 20 ads, and viewed the entire set of 100 ads over five sessions (some users viewed slightly lesser ads). Upon viewing each ad, the raters had a maximum of 10 seconds to input their engagement scores via mouse clicks using a prompt shown in Fig~\ref{eng_scale}. The Emotiv device comprises of 14 electrodes (locations shown in Fig~\ref{eeg_locs}), and has a sampling rate of 128 Hz. Upon experiment completion, the EEG recordings were segmented into \textit{epochs}, with each epoch denoting the viewing of a particular ad. Upon removal of noisy epochs. We recorded a total of 1738 epochs. Each ad was preceded by a 1s fixation cross to orient user attention, and to measure resting state EEG power used for baseline power subtraction. The EEG signal was band-limited between 0.1--45 Hz, and independent component analysis (ICA) was performed to remove artifacts relating to eye movements, eye blinks and muscle movements. The following section describes the techniques employed for content-centered AR and user-centered AR.

\section{Experiments and Results}\label{ER}

\begin{table*}[htbp]

\fontsize{10}{10}\selectfont
\renewcommand{\arraystretch}{2}
\centering
\caption{Advertisement Engagement Prediction from EEG analysis. F1 scores are presented in the form $\mu \pm \sigma$. } \label{tab:uscap}
\begin{tabular}{|c|c|c|c|}
  \hline
	\multicolumn{1}{|c|}{\textbf{Method}} & \multicolumn{3}{c|}{\textbf{Engagement}} \\ \hline 
	\multicolumn{1}{|c|}{~} & {\textbf{F1 for first 30 s (F30)}} & {\textbf{F1 for last 30 s (L30)}} & {\textbf{F1 for last 10 s (L10)}}  \\ \hline

	
	\textbf{LDA} & 0.6954 $\pm$ 0.0207 & 0.6781 $\pm$ 0.0256 & 0.6644 $\pm$ 0.0222 \\
	\textbf{LSVM} & 0.7018 $\pm$ 0.0201 & 0.6888 $\pm$ 0.0217 & 0.6647 $\pm$ 0.0211 \\
	\textbf{RSVM} & \textbf{{0.7091 $\pm$ 0.0224}} & \textbf{0.6803 $\pm$ 0.0204} & \textbf{{0.6732 $\pm$ 0.0263}} \\ \hline
\end{tabular}
\end{table*}

We first provide a brief description of the classifiers used and settings employed for engagement estimation, where the objective is to assign a binary (high/low) label for engagement evoked by each ad, using the extracted EEG  features. The ground truth here is provided by the mean of annotator ratings for each ad followed by thresholding by the grand mean of all ad ratings. Experimental results will be discussed hereafter.

\paragraph*{Classifiers:} We employed the Linear Discriminant Analysis (LDA), linear SVM (LSVM) and Radial Basis SVM (RSVM) classifiers in our AR experiments. LDA and LSVM separate high/low labeled training data with a hyperplane, while RSVM is a non-linear classifier which separates high and low classes, linearly inseparable in the input space, via transformation onto a high-dimensional feature space. We notice that \textit{RSVM} classifier produces the best F1-scores for engagement.

\paragraph*{Metrics and Experimental Settings:} We used the F1-score (F1), defined as the harmonic average of the precision and recall as our performance metric, due to the unbalanced distribution of positive and negative samples.

The 1738 clean epochs obtained from the EEG was used for user-centered analysis. To maintain dimensional consistency for subsequent principal component analysis (PCA), we performed user-centric AR experiments with (a) the first 3667 samples (30s of EEG data), (b) the last 3667 samples (30s of EEG data) and (c) the last 1280 samples (10s of EEG data) from each epoch. Each epoch sample comprises data from 14 EEG channels, and the epoch samples were input to the classifier upon vectorization and dimensionality reduction of the time-domain EEG data by PCA.

In this work, we use only time-domain EEG information. As we evaluate engagement performance on a small dataset, AR results obtained over 10 repetitions of 5-fold cross validation (CV) (total of 50 runs) are presented. CV is typically used to overcome the \textit{overfitting} problem on small datasets, and the optimal SVM parameters are determined from the range $[10^{-3},10^{3}]$ via an inner five-fold CV on the training set. Finally, in order to examine the temporal variance in AR performance, we present F1-scores obtained over (a) first 30s (F30), (b) last 30s (L30) and (c) last 10s (L10). These settings were chosen to study the efficiency with which engagement prediction can be realized at different temporal segments of the video to study whether there is a potential accuracy drop-off over time.

\subsection{Discussion}
Table \ref{tab:uscap} summarizes the results that we get for the various experiments for engagement estimation. The best performance that we achieve among the tested methods is using the RSVM classifier for the first 30 seconds of the ads (F1 = 0.709). RSVM is also uniformly the best classifier as compared to both LDA and LSVM, which points to the possibility that the decision boundary for engagement related statistics in EEG data is non-linear.

From the observations we infer that engagement recognition is lower in the L30 and L10 conditions, with a clear drop in accuracy later in the video. This could be because of a drop in engagement levels towards the end of the advertisement where strictly product related information is more prevalent as opposed to an engaging storyline in the beginning.

The observations also highlight the limitation of using a \textit{single} engagement label for the whole video (as opposed to dynamic labeling).

\section{Conclusion and Future Work}\label{CFW}
This work presents a first step towards using EEG data to directly predict the engagement level of a user while watching a video advertisement. There certainly exist many limitations that can be worked on in the future. Some of these are:-

\begin{itemize}
\item The level of engagement is treated here as a binary high or low label only. A much better and more rigorous way of doing this would be to annotate a continuous level of emotion at every time point using an annotation tool like Feeltrace \cite{cowie2000feeltrace} and a regression to estimate the continuous value instead of a categorical classification. However, such an annotation is very time consuming and has issues to deal with like synchronization.
\item This study only deals with the EEG modality. Combining it with other modalities that could be useful in a similar environment such as facial expressions, eye tracking, GSR, or even audiovisual content analysis would be an interesting avenue to explore.
\item This study only considers conventional classifiers (LDA and SVM). It would be interesting to see how the recent advances in deep learning (especially using transfer learning for small dataset sizes) can work on such a problem.
\item This study only considers the raw time domain EEG data that is cleaned using ICA and subsequently vectorized, dimensionality reduced, and then passed to a classifier. A lot of EEG related literature suggests that spectral features from the alpha, beta, gamma and theta bands are useful for EEG classification. However, these conventional features did not work as well for us and the time domain summary statistics using PCA gave us the best performance.
\end{itemize}

However, despite these limitations, the findings of this study offer an important insight into how a portable and commercially viable EEG device may be used for engagement measurement in the wild. This can have important industrial applications in the field of ad impact analysis. There exist modern tools by companies like Realeyes \cite{realeyes} such as \textit{Creative Testing} that provide second-by-second insights into advertisement engagement levels. They are based on principles of implicit tagging \cite{pantic2009implicit,soleymani2012mahnob} that are based on the user not having to consciously annotate any content in the video but their implicit responses (via EEG in our case) being able to automatically deduce the required value. The low cost and portability of the EEG headset used in the study further illustrates that the prototypical engagement estimation system can be subsequently developed into a viable commercial product.


\bibliographystyle{ACM-Reference-Format}
\bibliography{engagement_icmi_workshop} 

\end{document}